# WST, the Wide-field Spectroscopic Telescope: Telescope structure FE analyses

Simone D'Auria[a*], Vincenzo Cianniello[b], Ciro Del Vecchio[a], Vincenzo De Caprio[b], Philippe Dierickx[c], Gaston Gausachs[d], Tony Trauvillion[d], William Sutherland[e], Olga Bellido[f]

[a]INAF-Osservatorio Astrofisico di Arcetri, Largo Enrico Fermi 5, 50125 Firenze, Italy,

[b]INAF-Osservatorio Astronomico di Capodimonete, Salita Moiariello 16, 80131 Napoli, Italy

[c]Centre de Recherche Astrophysique de Lyon, France, [d]Australian National University, Australia

[e]Queen Mary Univeristy of London, [f]Leibniz-Institut für Astrophysik Potsdam, Germany

*simone.dauria@inaf.it

## ABSTRACT

The altitude structure of the Wide-field Spectroscopic Telescope (WST) is designed to support and position both primary and secondary mirrors, and it will be made of structural steel. Due to its dimensions and functions, the altitude structure is a substantial part of the WST facility, and its weight, performance, and dynamic behavior play a critical role in the functioning of the telescope. This paper discusses the iterative process starting from the preliminary structural layout and leading to the optimization of the entire structure. A Finite Element (FE) model was developed, defining the detailed dimensions and cross-sections of each assembly's beams and plates under representative boundary conditions, in order to correctly simulate the operational environment. This model enables accurate estimation of structural weight, mechanical deformations, and stresses, as well as its frequency response for the evaluation of both local and global resonance modes. Based on the initial results, obtained using the preliminary beam cross sections and shell thickness, several assumptions were formulated to drive the mechanical optimization of the altitude structure. The outcome of this work consists of a refined structural configuration and the formulation of the governing design criteria.

## 1. INTRODUCTION

The Wide-field Spectroscopic Telescope (WST) [1] [2] is an innovative next-generation facility designed to enable breakthrough science across several areas of modern astrophysics. WST is conceived as a 12-meter-class wide-field survey telescope capable of simultaneous Multi-Object Spectroscopy [3] (MOS), High Resolution [4] and Low Resolution [5], and Integral Field Spectroscopy (IFS) [6] observations. Within the Telescope structure and dome design [7], the Altitude Structure represents one of the main load-bearing subsystems, mainly supporting and positioning both the primary and the secondary mirror. Due to its dimensions and structural role, its mechanical behavior strongly affects the overall performance of the telescope. This work presents a preliminary structural characterization of the Altitude Structure through the development of a Finite Element (FE) model and an initial optimization study. The analysis aims to investigate the main static and dynamic characteristics of the Structure, including its stiffness, stress distribution, and dynamic response, to identify the most critical aspects of the current design and guide subsequent development stages.

## 2. ALTITUDE STRUCTURE DESCRIPTION

The Altitude Structure of the WST represents the main load-bearing subsystem of the Telescope. It is responsible for supporting both primary and secondary mirrors, together with the central optical tower hosting part of the optical chain and instrumentation (field corrector and atmosphere dispersion compensator, MOS focal plane, M3, M4, etc). Due to its overall dimensions and mass, this Structure has a key role in the determination of the global stiffness and dynamic response of the Telescope assembly. The overall envelope dimensions of the WST Altitude Structure are approximately 15.5 meters in width, 16.5 meters in height, and 14.5 meters in depth, making it one of the main structural subsystems of the telescope facility.

The preliminary structure layout was provided as a geometric wireframe model outlining the main beam paths and node locations. This model included the primary mirror (M1) supporting truss, the secondary mirror (M2) support structure, and the central tower connected to the optical path assemblies and focal instrumentation. Moreover, this assembly is supported by two lateral C-rings, which interface with the telescope fixed structure and define the altitude rotation axis. These components, provided as solid bodies, provide the mechanical interface between the moving part of the Telescope and the Nasmyth platform assembly.

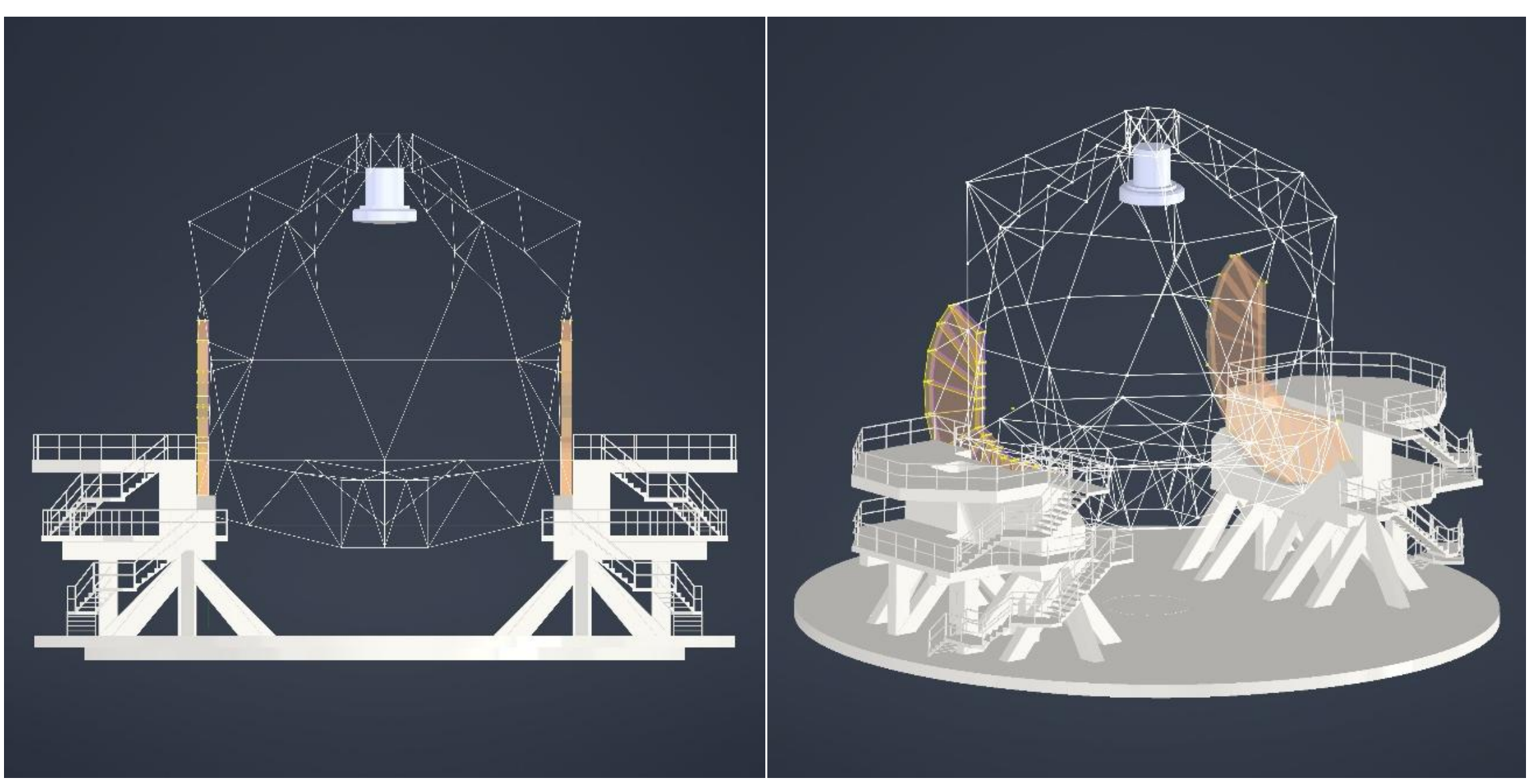

*Figure 1a, 1b - CAD model of the Telescope including the wireframe of the Altitude Structure, Nasmyth platforms and rotating platform (grey), C-rings (orange) and M2 (light blue)*

From a mechanical modelling perspective, the Altitude Structure can be divided into two main substructures: a beam-base truss assembly and the two lateral C-rings modeled as continuous shell structures.

## 3. FINITE ELEMENT ANALYSES

A Finite Element Model (FEM) of the WST Altitude Structure was developed to perform a preliminary assessment of the subsystem's static and dynamic behavior. The model was created starting from the conceptual structural layout described in the previous chapter, to preserve the main load paths and global stiffness characteristics while maintaining a proper computationally efficient representation for iterative design studies [8]. The computational tool used for these studies was COMSOL Multiphysics® [8]. The structural discretization was based on the use of beam and shell finite elements, according to their geometrical and mechanical characteristics. This approach enabled a direct assignment of the preliminary cross-sectional properties of the beam elements and an easy specification for the thickness of the shell elements [9]. Accordingly, the C-rings, initially provided as a solid geometry, were converted into geometric surfaces at the mid-thickness plane. This modeling strategy allowed relatively rapid analyses without losing the accuracy of the results.

### 3.1 MODELING ASSUMPTIONS

Several simplifying assumptions were introduced in the FE Model to analyze the global structural behavior of the Altitude Structure within a reasonable range of hypotheses. All beam elements were assumed to be rigidly connected at the corresponding nodes. Beam cross-sections were initially assigned using standard structural steel profiles selected by the estimated load distribution and structural functions. At the same time, a constant shell thickness value was adopted following the dimensions provided by the solid elements of the Altitude Structure's original design. The connections between the beam elements, through their common nodes, assumed a perfectly rigid connection node-to-node. That meant that the common nodes at the ends of two different beam elements were set to have the same deformation and rotation field. In this way, the common nodes transmitted axial force, shear, bending moment, and torsion. From a stiffness point of view, this allowed the Altitude Structure to be considered equivalent to a single continuous structure assembly.

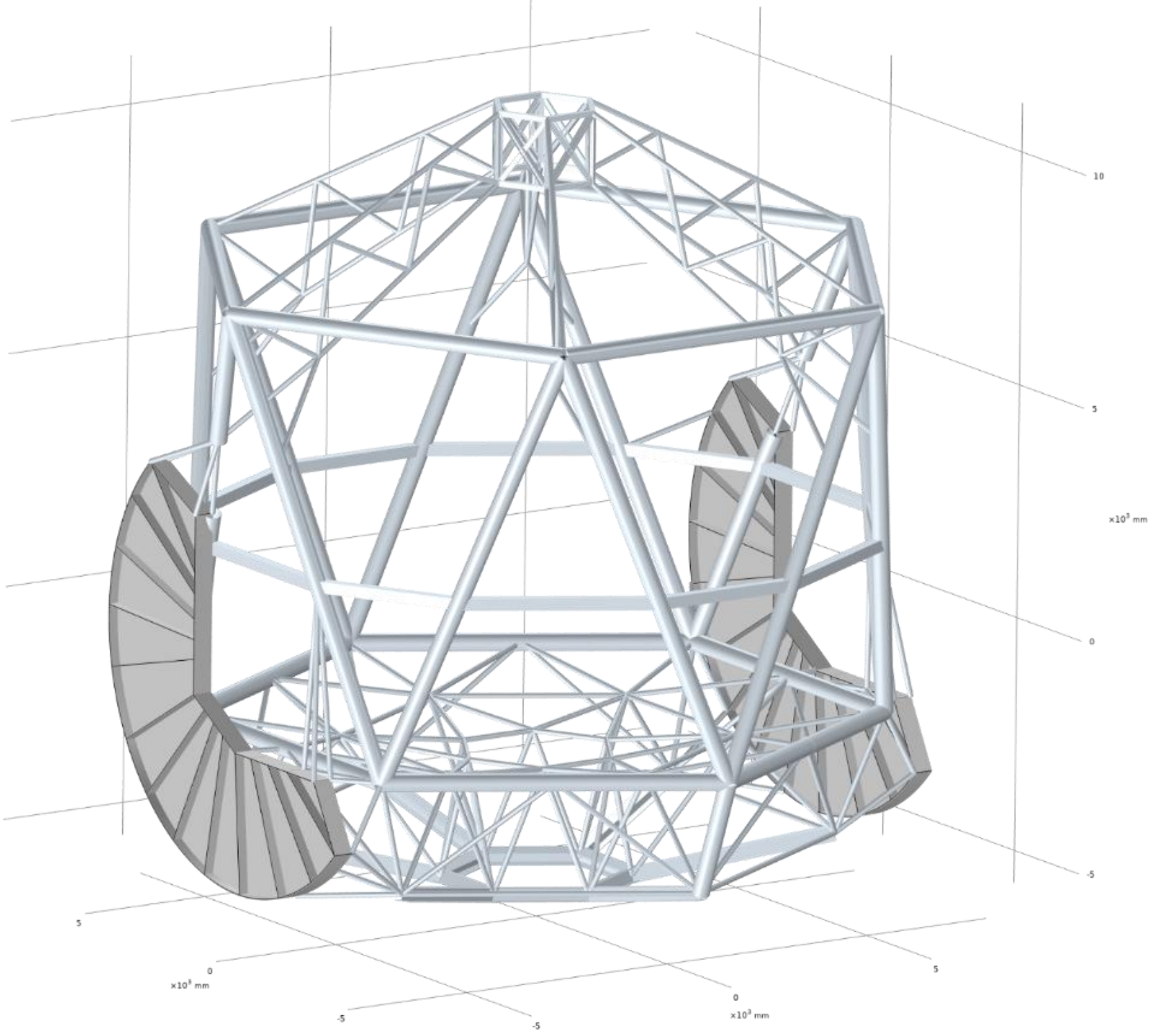

*Figure 2 - Altitude Structure represented with the detailed cross-section selected during the first draft of the model*

Beam and shell components were coupled through the Beam-Shell Connection formulation implemented in COMSOL Multiphysics. The connection ensured kinematic continuity between the two element types by transferring both translational and rotational degrees of freedom, thereby allowing a realistic representation of the interaction between the truss members and the C-ring assemblies. This simplified approach was considered adequate for the preliminary global analyses presented in this work. The material properties of all structural components were assumed to correspond to standard structural steel. Linear elastic behavior was considered throughout the analysis.

Non-structural components, as well as secondary mechanical interfaces and local mounting connections, were not explicitly modeled at this stage. Their mass contribution was included only when relevant to the global dynamic response and the static deflection under gravitational load.

### 3.2 BOUNDARY CONDITIONS AND LOAD CASES

The boundary conditions of this model were selected to reproduce the operational constraints of the Telescope Altitude Structure. The interface between the C-rings and the fixed Telescope structure was modeled using a simplified yet intentionally conservative representation, given the absence of a fully detailed definition of the mechanical interface at this design phase. This was a crucial aspect of the model setup as it governed the realistic global stiffness contribution of the whole Altitude Structure. For this reason, the connection between the Altitude Structure and the stationary support was idealized by constraining two representative edges of the shell elements (one for each C-ring). At this stage, a fixed boundary condition was applied to these edges to approximate the effects of the Altitude drive system in a locked configuration, which was considered a reasonable assumption. This approach was consistent with the common practice in a preliminary analysis of such a large Telescope in which the bearing-drive interfaces are often represented through fixed constraints. This assumption provided a quite conservative estimate of the structure stiffness and a slight overestimation of the system eigenfrequencies, which remained acceptable for an early-stage design screening and detection of potential dynamic criticalities.

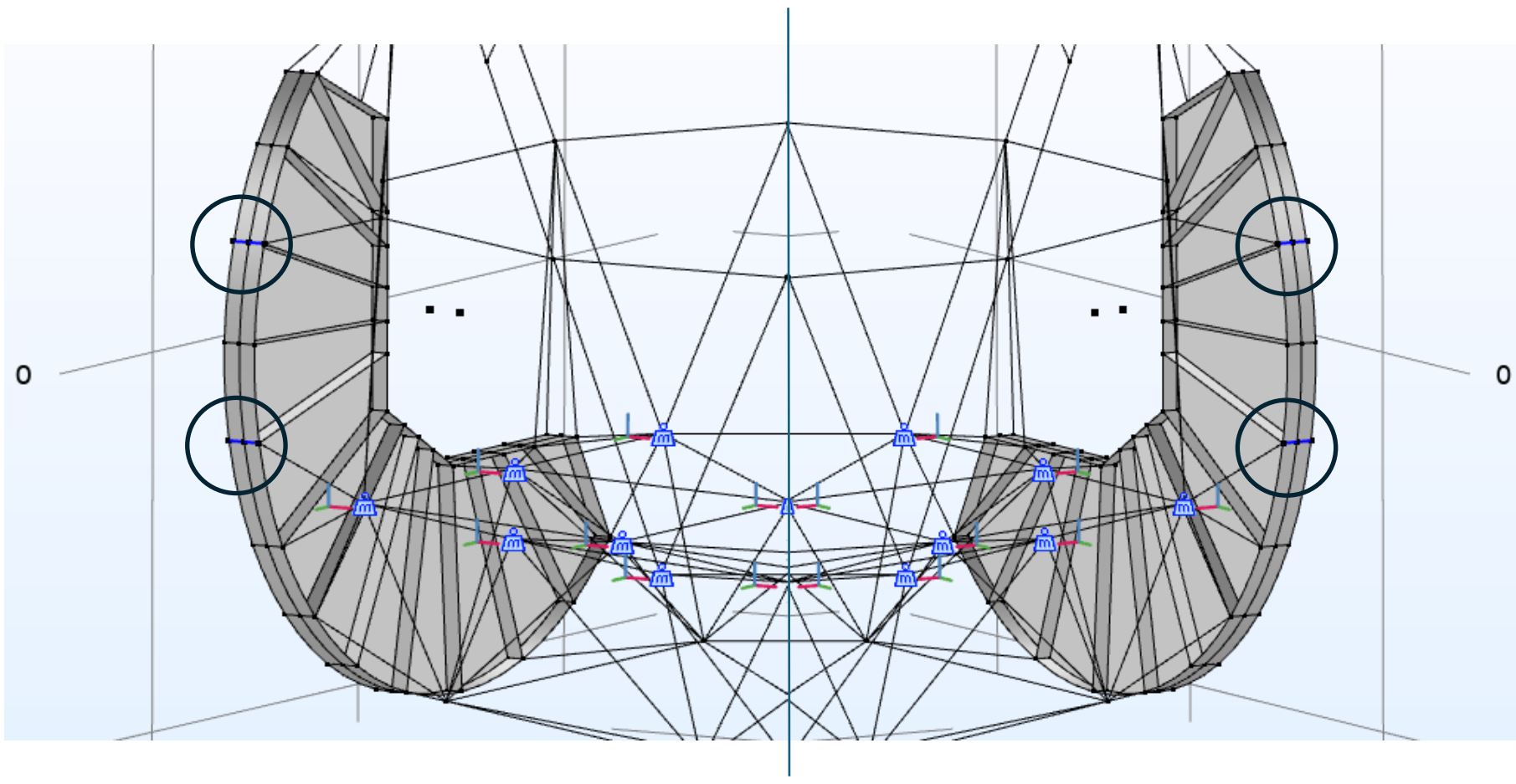

*Figure - 3a, 3b - Fixed boundary conditions (in circles). This selection changes while changing the altitude angle of the Telescope and the hypothetical angle between the rotors*

### 3.3 REPRESENTATION OF THE OPTICAL COMPONENTS

The primary and secondary mirrors were not explicitly modeled through detailed geometrical representations. Instead, their contribution to the structural behavior of the telescope was included by means of point mass elements applied at the relevant interface locations. For the primary mirror (M1), the total mirror mass was distributed among the beam nodes defining the primary mirror support structure. The corresponding point mass elements were assigned to these nodes to reproduce the gravitational and inertial effects of M1 while maintaining a computationally efficient model. A different approach was adopted for the secondary mirror (M2). Owing to its position at the end of the telescope upper structure and its suspension below the spider assembly, its influence on the dynamic response could not be adequately represented through translational masses alone. Therefore, M2 was modeled using point-mass elements combined with equivalent rotational inertias applied at the corresponding support nodes. The rotational inertia was estimated by approximating the secondary mirror assembly as a solid cylinder with equivalent mass properties. The resulting inertia tensor was assigned to the point mass elements, thereby allowing the model to account for the rotational effects associated with the suspended mass and providing a more representative description of the dynamic behavior of the upper telescope assembly.

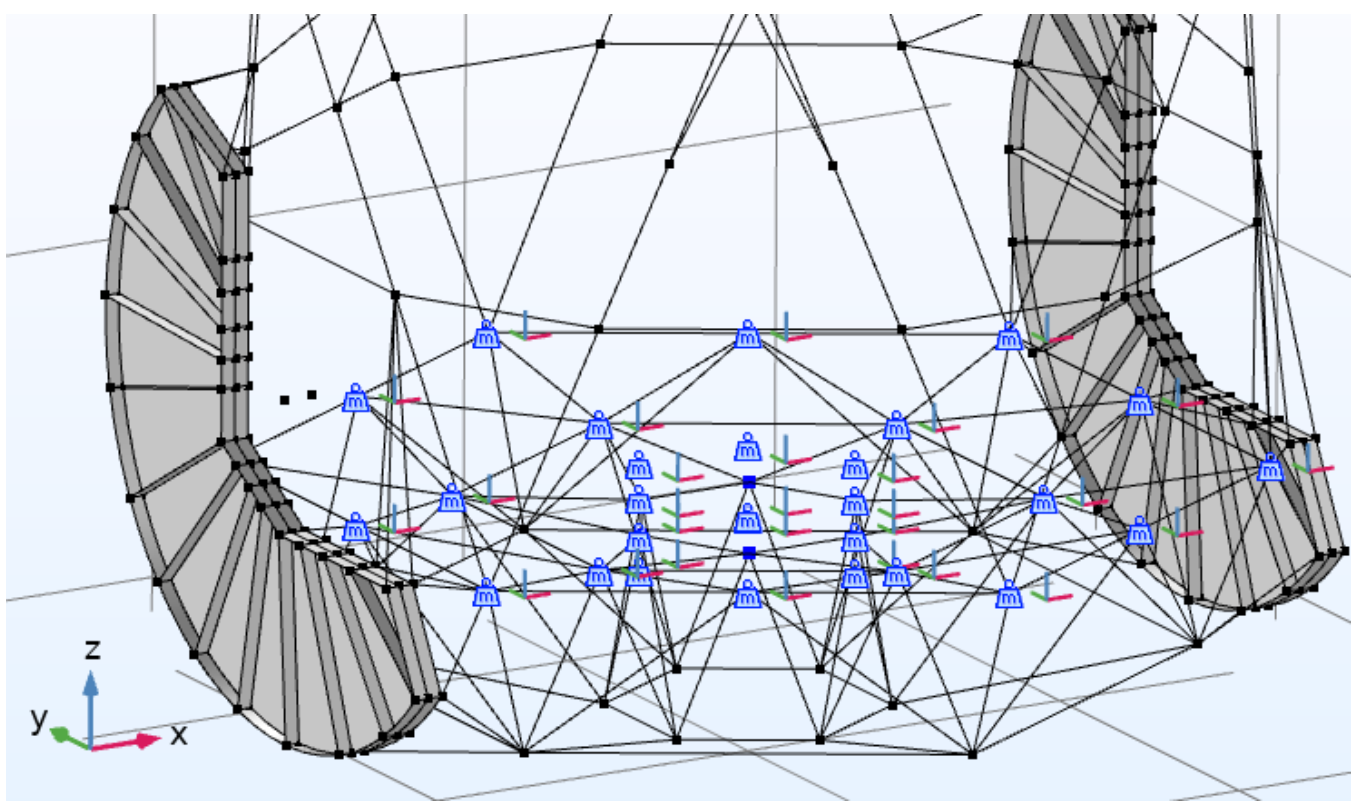

*Figure 4 - Mass points (blue) that replicate the M1 distribution on the wireframe*

The adopted modeling strategy enabled the dominant inertial contributions of the optical components to be captured without introducing additional geometrical complexity, making the approach particularly suitable for the preliminary structural characterization and optimization activities presented in this work.

# 4. RESULTS

## 4.1 STATIC STRESS AND DEFORMATIONS

A preliminary static assessment of the altitude structure was performed to evaluate the global stress distribution and deformation patterns under operational gravitational loading. The analyses were repeated for different telescope altitude angles, corresponding to varying gravity directions with respect to the structure, and for different locked-rotor configurations of the altitude driving system, represented through alternative constraint conditions at the C-ring interfaces. The obtained stress distributions did not reveal significant structural criticalities. Maximum stress values were found to be of the order of a few hundred MPa, remaining compatible with the use of structural steel as the reference material. Furthermore, the highest stress concentrations were localized in limited regions, primarily at the ends of beam elements and in the transition areas between the beam-based truss structure and the shell-modelled C-rings. These localized peaks were strongly influenced by the simplified representation adopted for the beam-shell connections and should therefore be interpreted with caution. A more detailed assessment of these regions will be carried out during subsequent design stages, when the geometry and mechanical interfaces of the connection assemblies will be available.

The overall deformation levels were found to be limited for all the investigated configurations. Attention was devoted to the relative displacement between the primary mirror (M1) and the secondary mirror (M2), as this parameter directly affected the optical alignment of the telescope. The relative displacement was estimated by comparing the average displacement of the nodes associated with the M1 and M2 support regions. For the most unfavorable loading configuration, the relative displacement between the two optical elements remained less than a millimeter. Both the axial component, associated with defocus effects, and the transverse component, associated with decentering and alignment errors, remained within a range that did not indicate major structural concerns at this stage of the design.

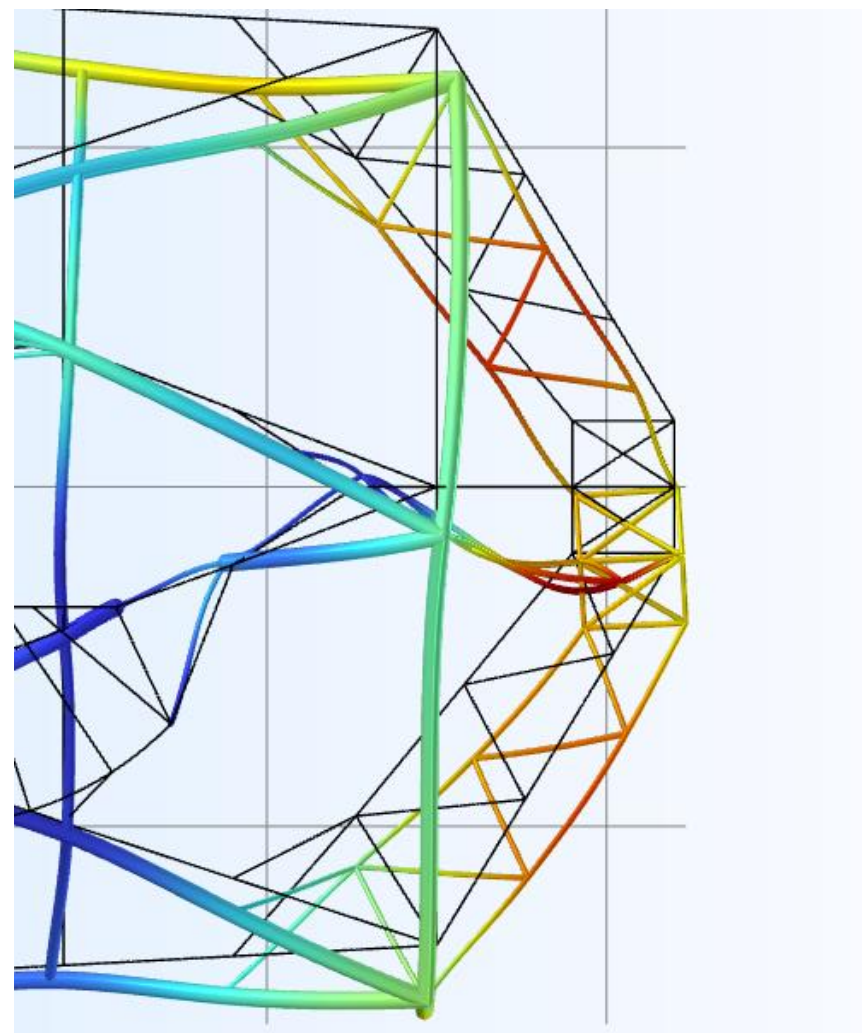

*Figure 5 - Static deformation (colored) with respect to the undeformed structure (black lines) due to the gravity effect with the Telescope pointing at the horizon*

The static analyses confirm that the preliminary structural layout provided adequate global stiffness and did not exhibit critical stress or deformation levels, supporting the continuation of the study towards the characterization and optimization of the dynamic behavior of the altitude structure.

### 4.2 DYNAMIC RESPONSE

Following the definition of the boundary conditions described in Section 3, a modal analysis was performed to characterize the dynamic behavior of the altitude structure and identify the dominant vibration modes of the telescope assembly. The first eigenmode was associated with the spider assembly supporting the secondary mirror (M2). Due to the cantilevered configuration of the upper structure and the position of M2 at the maximum distance from the constrained regions located at the C-rings, this mode was characterized by a rotational motion of the spider around its attachment points. Although this mode exhibited the lowest natural frequency of the structure, its effect on the relative position of the primary and secondary mirrors was limited, and therefore it is not expected to represent a major concern for the overall optical performance of the Telescope. It should also be noted that the spider geometry was intentionally maintained centrosymmetrical with respect to the primary mirror segmentation pattern, as alternative symmetry configurations could complicate segment phasing calibration and alignment procedures.

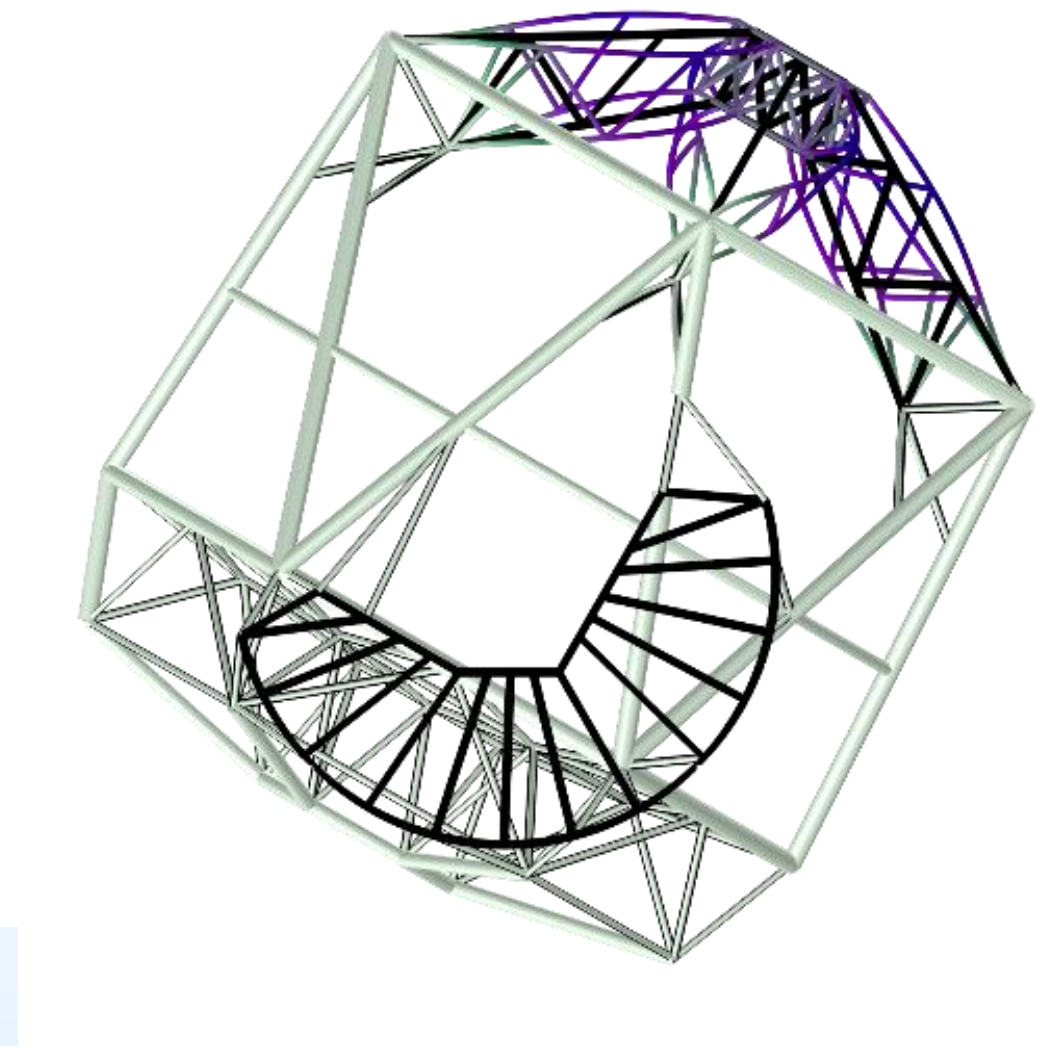

*Figure 6 - First eigenfrequency of the Structure, involving the spider structure that mounts M2*

The second mode corresponded to the first global mode of the altitude structure under locked-rotor conditions. In this configuration, the altitude drive system was assumed to be mechanically locked, and the mode was characterized by a rotation of the entire structure about the telescope x-axis. Similar modes are commonly observed in large telescope structures and are often used as a reference indicator of the global stiffness of the altitude assembly and its interfaces with the supporting structure. The associated natural frequency showed a moderate dependence on the telescope altitude angle. As the telescope approached the horizon position, the frequency gradually decreased and reached values of approximately 8 Hz. This behavior reflects the variation of the gravity-induced preload and the corresponding redistribution of stiffness within the structure. The third mode was characterized by a lateral displacement of the altitude assembly, predominantly along the x-direction. This mode involved the global deformation of the main structural framework and exhibited natural frequencies comparable to those of the locked-rotor mode, particularly at low altitude angles.

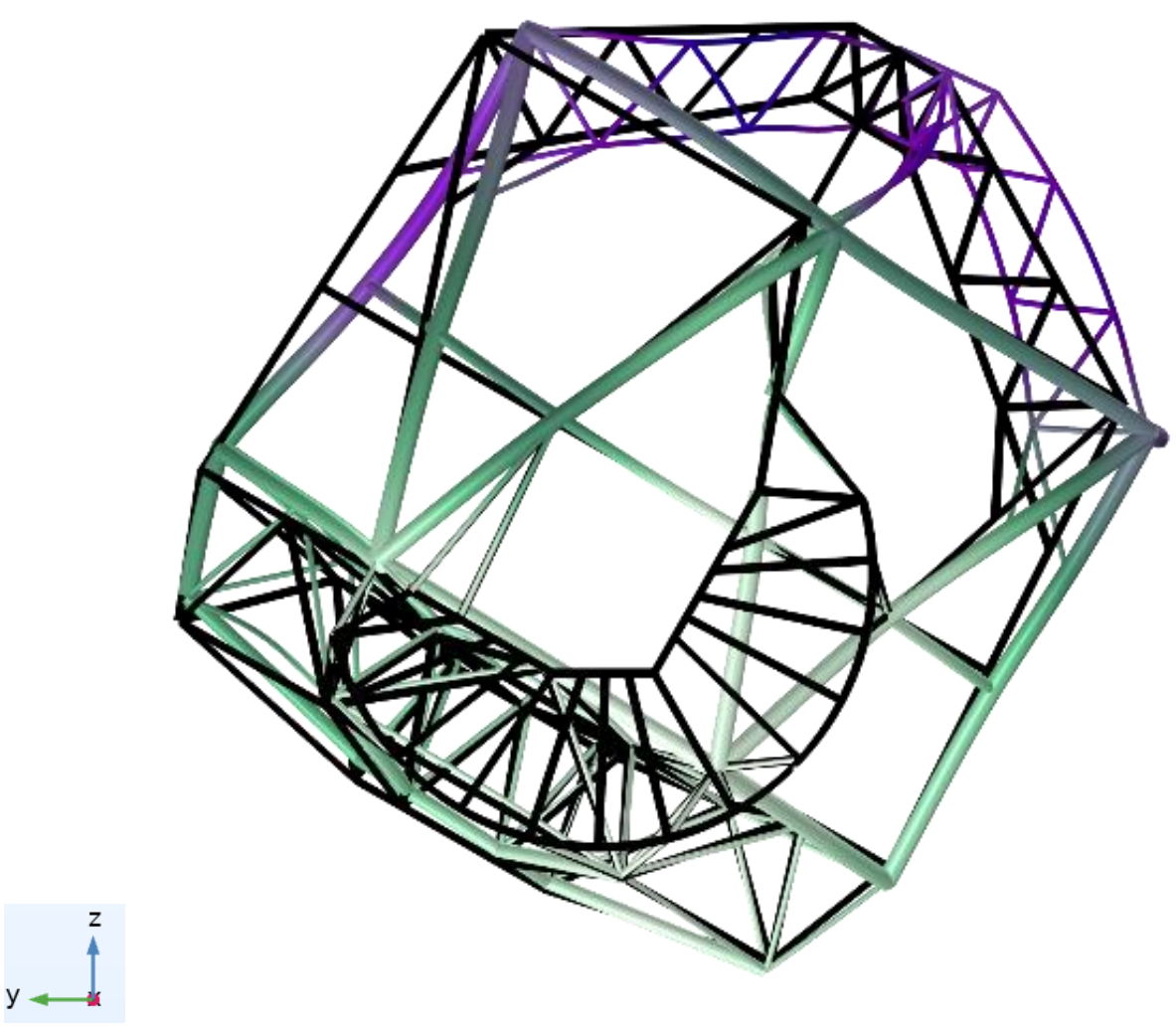

*Figure 7 - Second eigenfrequency of the Structure: Locked rotor mode. Involving the whole structure rotation around the x-axis*

The proximity between the frequencies of the second and third modes indicates that the global dynamic response of the structure is governed by the overall stiffness of the altitude assembly rather than by localized structural components. Consequently, these modes were identified as the primary targets of the subsequent optimization activities aimed at increasing the natural frequencies of the telescope structure and improving its dynamic performance.

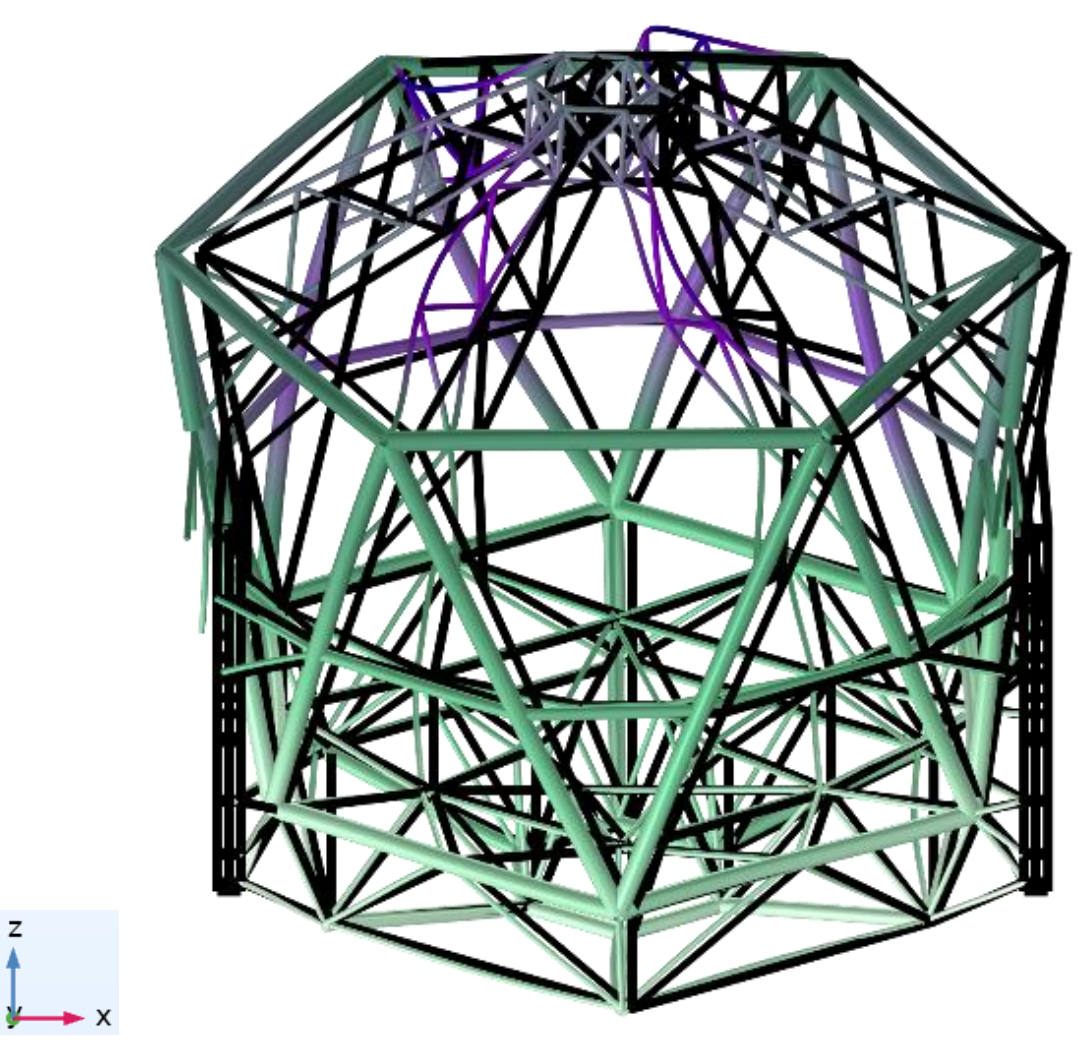

*Figure 8 - Third eigenfrequency of the Structure, lateral shift along the x-axis*

### 4.3 OPTIMIZATION

The optimization activity focused on a limited set of design variables representative of the main structural components of the altitude assembly. A total of six parameters were considered, including the outer diameter and wall thickness of selected groups of beam elements, together with the shell thickness assigned to the C-rings. The parameters were chosen to allow modifications of the global stiffness distribution while preserving the overall structural layout and load paths defined in the preliminary design.

The objective function was defined as the sum of the natural frequencies associated with the second and third vibration modes, divided by the total mass of the whole structure:

$$J = \frac{f_2 + f_3}{m}$$

thereby promoting simultaneous improvement of both global dynamic characteristics. This formulation was commonly adopted in preliminary structural optimization studies when multiple low-frequency modes contribute to the overall dynamic performance of the system, and no single mode can be identified as uniquely critical. Maximizing the combined frequency value encouraged a more balanced stiffness increase throughout the structure and reduced the risk of improving one mode at the expense of another. At the same time, a mass constraint was imposed to ensure that the total structural mass remained below 150 tons, preventing excessive stiffness increases at the expense of overall weight.

In addition to the objective function, a mass constraint was introduced to ensure compatibility with the overall telescope design requirements. The total structural mass of the altitude assembly was therefore constrained not to exceed 150 tons throughout the optimization process. This limitation reflects the practical need to balance dynamic performance against manufacturing, integration, and operational considerations.

The optimization results showed that the most effective design modifications consisted of increasing the stiffness of specific beam groups participating in the global deformation modes, together with a moderate increase in the thickness of the C-ring assemblies. These changes produced a measurable upward shift of both target frequencies while maintaining the total structural mass within the prescribed limit.

The optimized configuration obtained from this preliminary study provides an improved baseline for future design iterations and confirms the effectiveness of targeted stiffness redistribution as a means of enhancing the dynamic performance of the altitude structure.

## 5. CONCLUSION

A preliminary finite element model of the WST altitude structure has been developed to assess the static and dynamic behavior of the subsystem during the early design phases. Starting from the conceptual structural layout, a simplified beam-and-shell representation was implemented, allowing the main load paths, stiffness characteristics, and inertial contributions of the optical assemblies to be captured with limited computational effort.

The static analyses performed under different telescope altitude angles and locked-rotor configurations did not reveal significant structural concerns. Stress levels remained compatible with the use of structural steel, while the observed stress concentrations were mainly confined to localized beam-shell transition regions that will be further refined in future design stages. Similarly, the predicted deformations and relative displacements between the primary and secondary mirrors were found to be limited and indicated adequate global stiffness of the preliminary structural configuration.

The modal analysis enabled the identification of the dominant vibration modes of the altitude assembly. In particular, the global locked-rotor and lateral deformation modes were recognized as the most relevant dynamic features of the structure and were therefore selected as the target of a preliminary optimization activity.

A parametric optimization study was subsequently carried out by varying the geometric properties of selected beam groups and the thickness of the C-ring assemblies while enforcing a maximum structural mass of 150 tons. The results demonstrate that a targeted redistribution of stiffness can effectively increase the natural frequencies associated with the critical global modes, providing a more favorable dynamic behavior without significant impact on the overall structural concept.

Future activities will focus on the refinement of the finite element model through the introduction of more detailed interface representations, improved characterization of the altitude drive and bearing assemblies, and the inclusion of additional subsystem masses as their design matures. The optimized configuration presented in this work will serve as the baseline for subsequent structural developments and for more advanced analyses aimed at supporting the final design of the WST altitude structure.

## ACKNOWLEDGMENTS

This project has received funding from the European Union’s Horizon Europe research and innovation programme under grant agreement No. 101183153.

## REFERENCES

[1] «WST official website,» [Online]. Available: https://wstelescope.eu/.

[2] Bacon R. et al, "WST - Widefield Spectroscopic Telescope: the next leap in wide-field," this conference, 2026.

[3] Lee D. et al, «WST, the Wide-field Spectroscopic Telescope: progress on the design of the instruments,» this conference, 2026.

[4] Tozzi A. et al, «WST, the Wide-field Spectroscopic Telescope: the MOS-HR spectrograph module,» this conference, 2026.

[5] Buffat D. et al, «WST, the Wide-field Spectroscopic Telescope: design trade-offs for the low-resolution multi-object spectrograph instrument,» this conference, 2026.

[6] Cudennec C. et al, «Guiding Design Choices for Wide-Field IFS: Trade-Offs Between Replication and,» this conference, 2026.

[7] Gasusachs G. et al, «The Wide-field Spectroscopic Telescope (WST): telescope structure and enclosure design,» this conference, 2026.

[8] Cianniello V. et al, «FE model evolution of the Main Support Structure interfaces to the ELT Nasmyth Platform from PDR to FDR,» SPIE, n. 13096, 2024.

[9] COMSOL Multiphysics v. 6.4, [Online]. Available: www.comsol.com.

[10] Cianniello V. et al, «Finite element modeling technique: trade-off between two different FE models of a mechanical selector for astronomical instrumentation,» SPIE, n. 12184, 2022.